\documentclass[a4paper,11pt]{article}
\usepackage{graphicx,amssymb,bm,latexsym,color,epsf}
\makeatletter
\@addtoreset{equation}{section}

\makeatother
\newcommand{\be}{\begin{equation}}
\newcommand{\ee}{\end{equation}}
\newcommand{\bea}{\begin{eqnarray}}
\newcommand{\eea}{\end{eqnarray}}
\newcommand{\vs}[1]{\vspace{#1 mm}}
\newcommand{\hs}[1]{\hspace{#1 mm}}

\newcommand{\s}{\sigma}

\newcommand{\pa}{\partial}

\newcommand{\nn}{\nonumber\\}
\newcommand{\p}[1]{(\ref{#1})}

\newcommand{\cD}{{\cal D}}

\newcommand{\cL}{{\cal L}}
\newcommand{\br}{\bar R}
\newcommand{\bR}{\bar R}
\newcommand{\bg}{\bar g}

\newcommand{\bDelta}{\bar\Delta}

\newcommand{\bnabla}{\bar\nabla}

\newcommand{\Det}{{\rm Det}}

\newcommand{\diff}{\mathit{Diff}}
\newcommand{\sdiff}{\mathit{SDiff}}
\newcommand{\lich}{{\Delta_L}}

\begin{document}
\begin{flushright}
KU-TP 070 \\
\today
\end{flushright}

\begin{center}
{\Large\bf The path integral of unimodular gravity}
\vs{10}

{\large
R. de Le\'on Ard\'on$^{a,b,}$\footnote{e-mail address: rdeleon@sissa.it}
N. Ohta,$^{c,d,}$\footnote{e-mail address: ohtan@phys.kindai.ac.jp}
and R. Percacci$^{a,b,}$\footnote{e-mail address: percacci@sissa.it}
} \\
\vs{10}
$^a${\em International School for Advanced Studies, via Bonomea 265, 34136 Trieste, Italy}

$^b${\em INFN, Sezione di Trieste, Italy}

$^c${\em Department of Physics, Kindai University,
Higashi-Osaka, Osaka 577-8502, Japan}

$^d${\em Maskawa Institute for Science and Culture,
Kyoto Sangyo University, Kyoto 603-8555, Japan}

\vs{10}
{\bf Abstract}
\end{center}
We compute the one-loop effective action in unimodular gravity,
starting from two different classical formulations of the theory.
We find that the effective action is the same in both cases, 
and agrees with the one of General Relativity.

\section{Introduction}

Unimodular Gravity (UG) is a reformulation of General Relativity (GR)
where the determinant of the metric is fixed a priori and
is not subjected to variation \cite{Anderson:1971pn,vanderBij:1981ym,Buchmuller:1988wx,
Ellis:2010uc,Henneaux:1989zc}.
The difference between these two theories is very subtle,
since in GR the determinant of the metric changes under diffeomorphisms
and in particular can be fixed locally to have a predetermined form.
One could thus see UG as a particular gauge-fixed version of GR.
Indeed, the equations of motion of the two theories are identical.
The only substantial difference, which is also the reason
for the recent interest in UG, is the status of 
the cosmological constant:
a coupling in the Lagrangian of GR, and an integration constant in UG.
As a consequence, it is expected that the problems of the
cosmological constant appears differently in the two theories,
especially when quantum effects are taken into account
\cite{Smolin,Padilla:2014yea,Bufalo:2015wda}.

At the classical level the two theories are ``almost'' equivalent,
in the sense that they only differ by the status of a single
global scaling mode of the metric \cite{Henneaux:1989zc}.
(We will not consider this important subtlety in this paper,
so by equivalence we will always mean ``almost equivalence''.)
The question then arises whether UG and GR are also 
equivalent at the quantum level.

As a concrete way of addressing this question,
we set out to straightforwardly compare the divergent 
part of the effective action for UG and GR,
both computed using the conventional one loop background field method.
It is known that the same coefficients appearing in the
logarithmically divergent part also appear in
certain finite nonlocal terms in the effective action.
Thus, different logarithmic divergences imply physical
inequivalence of the two theories.

There is no single unambiguous way to construct a quantum
theory from a classical one, and the answer may well depend on the
starting point and on details of the procedure followed.
The authors of (\cite{Padilla:2014yea,Bufalo:2015wda,Fiol:2008vk})
argue in favor of equivalence, possibly with qualifications.
On the other hand \cite{Alvarez}
undertook the calculation of logarithmic divergences in UG
and found a different result from GR.
This conclusion is also implicit in
\cite{Eichhorn,Benedetti:2015zsw,Saltas:2014cta}.

There are several ways of formulating UG.
We shall start from the most straightforward one,
which imposes the unimodularity constraint on the determinant 
of the metric.
Actually, given that we want to allow the metric to be dimensionful,
we cannot simply set $|g|=1$.
A more general statement is to fix a volume form 
\be
\omega\,\epsilon_{\mu_1\ldots\mu_d}\,dx^{\mu_1}\wedge\ldots\wedge dx^{\mu_d}\ ,
\ee
not related to the dynamical metric $g$, and to require
\be
\sqrt{|g|}=\omega\ .
\label{constraint}
\ee
The function $\omega$ is a scalar density of weight  one.
In the following we shall use the background field method
to construct the path integral for gravity,
and the background metric $\bg_{\mu\nu}$ will be assumed
to be unimodular in the same sense, namely
$\sqrt{|\bg|}=\omega$.

The action of Unimodular Gravity (UG) is
\be
S_{UG}(g)=Z_N\int d^dx\, \omega\, R
\qquad\mathrm{where}\qquad Z_N=\frac{1}{16\pi G}\ .
\label{action}
\ee
Since $\omega$ is a fixed non-dynamical background,
it breaks the diffeomorphism invariance 
to the group $\sdiff$ of volume-preserving,
or ``special'' diffeomorphisms.
\footnote{The relation between $\sdiff$
and $\diff$ parallels the relation between $SO(n)$ and $O(n)$.}
The infinitesimal generators of this group are vectorfields
$\epsilon^\mu$ satisfying 
\be
\nabla_\mu\epsilon^\mu=0\ .
\label{infgen}
\ee
Note that when one varies the action (\ref{action}), the term 
$g^{\mu\nu}\delta R_{\mu\nu}=\nabla_\mu\Omega^\mu$
can be discarded as in GR, since equation (\ref{constraint})
implies that $\nabla\omega=0$, and $\nabla$ can be integrated
by parts as usual. 
In particular, when 
$\delta g_{\mu\nu}=\nabla_\mu\epsilon_\nu+\nabla_\nu\epsilon_\mu$
invariance of the action under $\sdiff$
follows from integrations by parts, the Bianchi identity 
and (\ref{infgen}).
For discussions of more general $\sdiff$-invariant
theories of gravity see \cite{Alvarez:2006uu,Blas:2011ac,Bonifacio:2015rea}.

There exist local coordinate systems
where $\omega$ is constant.
Then, in the case when the coordinates are chosen to have dimension
of length, one can set $\omega=1$, and the metric would
be unimodular in the proper sense of the word.
We will stick to the standard terminology and call
UG a theory where (\ref{constraint}) holds,
even when $\omega$ is not a pure number and not constant.

The constraint (\ref{constraint}) could be imposed by 
adding to (\ref{action}) a term containing a Lagrange multiplier,
multiplied by $\sqrt{|g|}-\omega$.
This, however, would complicate the calculation of radiative corrections.
Since in practice the functional integration measure
is defined on the fluctuation field anyway,
a simpler alternative is to use an exponential parametrization
for the background field expansion
\cite{Kawai:1989yh,Eichhorn,pv1,nink}
\be
\label{exppar}
g_{\mu\nu}=\bg_{\mu\rho}(e^X)^\rho{}_\nu
\ee
and demand that $X^\rho{}_\nu=\bg^{\rho\sigma}h_{\sigma\nu}$ 
be traceless.
This automatically enforces unimodularity of $g_{\mu\nu}$,
and tracelessness is a linear condition that can
be straightforwardly imposed on the quantum field,
without using Lagrange multipliers.
We will refer to this formulation as ``minimal'' UG.

An alternative formulation consists in writing the
action in terms of an unrestricted dynamical metric 
$\gamma_{\mu\nu}$, conformally related to $g_{\mu\nu}$:
\be
g_{\mu\nu}=\gamma_{\mu\nu}\left(\frac{|\gamma|}{\omega^2}\right)^m\ .
\label{gamma}
\ee
Note that the fraction is a genuine scalar,
so that both $g$ and $\gamma$ are proper tensors,
and the standard formulae for the curvatures of conformally related metric can be applied.
For $m=-1/d$ the metric $g_{\mu\nu}$ automatically
satisfies the constraint (\ref{constraint}).
Then, we can take $\gamma$ as the fundamental dynamical variable
and write the action \cite{Alvarez:2006uu,Blas:2011ac}
\bea
S(\gamma)&=&S_{EH}(g[\gamma])
\nonumber\\
&=&Z_N\int d^dx\, |\gamma|^{1/d}\omega^\frac{d-2}{d}
\left[R[\gamma]+\frac{(d-1)(d-2)}{4d^2}
\left(|\gamma|^{-1}\nabla|\gamma|
-2\omega^{-1}\nabla\omega\right)^2\right]\ ,
\label{ester}
\eea
where $\nabla$ are the covariant derivatives constructed with
$\gamma_{\mu\nu}$.\footnote{Recall that under $\sdiff$ the determinant
$|g|$ is a scalar, so $\nabla_\mu|g|=\partial_\mu|g|$.}
This action is invariant under $\sdiff$ as well as an
additional Weyl group acting as
\be
\gamma_{\mu\nu}\to\Omega^2\gamma_{\mu\nu}\ ;
\qquad \omega\to\omega\ ,
\ee
which leaves the metric $g_{\mu\nu}$ invariant.

These two formulations differ in the way in which a scalar
degree of freedom is removed: either by imposing a constraint
or by the presence of a gauge freedom.
They are classically equivalent.
In this paper we compute the one-loop divergences of
UG in both formulations, 
and we shall see that the results are the same,
and also agree with those of GR.

We mention here another motivation for this calculation,
which actually prompted our initial interest in this problem.
In \cite{pereiraI} we computed the gravitational 
divergences of GR using as dynamical variable a
metric $\gamma_{\mu\nu}$ defined as in (\ref{gamma})
(with $\omega=1$) in the generic case $m\not=-1/d$.
(This was later generalized to higher-derivative gravity in \cite{pereiraII}.)
All the calculations were performed with a standard
gauge-fixing term depending on two parameters $\alpha$ and $\beta$.
Furthermore, the parametrization of the metric depended on
two parameters $m$ and $\varpi$ (the latter was
called $\omega$ in \cite{pereiraI,pereiraII}, but we use here
a different notation in order not to confuse it with the
fixed volume form $\omega$).
The results were found to be the same independently of
$\alpha$, $\beta$ and $\varpi$ in the limit $m\to-1/d$,
and independent of $\alpha$, $\beta$ and $m$ 
in the case when $\varpi=1/2$, which corresponds to using
the exponential parametrization for the metric.
In these two cases there was an enhanced gauge-independence.
However, the calculations in those papers did not cover the
unimodular case $m\to-1/d$, because our gauge fixing was 
not complete in that case due to the additional Weyl invariance.
There remained therefore the task of calculating the
divergences directly in the unimodular theory,
to see whether they agree with the limit $m\to-1/d$ or not.
This is what we will do here.

The calculation of the one-loop divergences requires that
we write the one-loop effective action as a path integral,
which can be expressed as a product of determinants.
In the minimal formulation,
a simple argument seems to suggest that the path integrals of
UG and GR should be different.
Classical UG is GR where we have partially
fixed the gauge by fixing the determinant of the metric
or, at the linearized level, we have set the trace of the
fluctuation to zero.
This partial gauge fixing requires a scalar ``ghost'' determinant \cite{pv1}.
On the other hand when we quantize UG the condition $h=0$ 
is already present at the classical level and there is no need to introduce a ghost.
The remaining gauge freedom can be fixed in the same way,
so there seems to be a net difference of a scalar determinant.
This argument is wrong, for reasons that will appear 
in section 3.2.
However, it points to subtleties in the definition
of the path integral.
For this reason we consider first the Hamiltonian 
definition of the path integral, which is more fundamental.
In section 2 we will calculate the path integrals of 
linearized GR and UG 
in certain gauges where they are explicitly seen to be the same.
It will be enough to do this on a flat background.
Unfortunately the Hamiltonian formalism is not well-suited
to discuss covariant gauges.
In section 3 we calculate the one-loop effective actions
of GR and UG by covariant methods.
The main point here will be identifying the proper way of
factoring the volume of the gauge group $\sdiff$.
We construct the proper gauge-fixed path integral for UG 
and find that it agrees, in a well-defined sense, 
with the one of GR.
Section 4 contains some comments and conclusions.

\section{Hamiltonian analysis of GR and minimal UG}

\subsection{Linearized GR}

In this section we prove, based on constrained
Hamiltonian formalism, that the
path integrals of GR and UG contain the same determinants.
Insofar as the Lagrangian path integral is derived from
the Hamiltonian one, this proof is somewhat more fundamental.
For our purpose it is going to be enough to consider the 
case of a flat background.
Then GR reduces to the Fierz-Pauli theory with Lagrangian density
\be
\label{fierzpauliaction}
\cL=
-\frac{1}{2}\partial_\alpha h_{\mu\nu}\partial^\alpha h^{\mu\nu}
+\partial_\alpha h_\mu{}^\alpha \partial_\beta h^{\mu\beta}
-\partial_\alpha h_\mu{}^\alpha \partial^\mu h
+\frac{1}{2}\partial_\alpha h \partial^\alpha h\ .
\ee
The canonical analysis of the Fierz-Pauli theory has been
discussed earlier in \cite{baaklini}.
Here we shall redo it in a different set of variables.

In view of the canonical analysis, we begin by decomposing 
all tensors into time (0) and space ($i,j\ldots$) components.
We rename the variables as follows:
$h_{00}=-2\phi$, $h_{0i}=v_i=v^T_i+\partial_i v$, where 
$\partial_i v^T_i=0$, and for the space metric we use
the York decomposition
\begin{equation}
h_{ij}=h^{\mathrm{TT}}_{ij}
+\partial_i\zeta_j+\partial_j\zeta_i
+\left(\partial_i\partial_j-\frac{1}{d-1}\delta_{ij}\partial^{2}\right)\tau
+\frac{1}{d-1}\delta_{ij}t\,,
\label{ap1}
\end{equation}
where $\partial_i\zeta_i=0$, $\partial_i h^{TT}_{ij}=0$,
$h^{TT}_{ii}=0$ (summed over $i$).
These are the variables that are often used in the analysis
of cosmological perturbations.

Under an infinitesimal gauge transformation 
$\epsilon_\mu=\{\epsilon_0,\epsilon_i\}$
\bea
\delta\phi&=&\dot\epsilon_0\\
\delta v_i&=&\partial_i\epsilon_0+\dot\epsilon_i\\
\delta h_{ij}&=&\partial_i\epsilon_j+\partial_j\epsilon_i
\eea
The transformation parameter can be decomposed in transverse and longitudinal parts
\be
\epsilon_i=\epsilon^T_i+\partial_i \epsilon
\ee
Then we get
\bea
\delta\phi&=&\dot\epsilon_0\\
\delta v&=&\epsilon_0+\dot\epsilon\\
\delta v^T_i&=&\dot\epsilon^T_0\\
\delta t&=&2\partial^2\epsilon\\
\delta\tau&=&2\epsilon\\
\delta \zeta_i&=&\epsilon^T_i\\
\delta h^{TT}_{ij}&=&0
\eea
From here we see that the scalar combinations:
\be
\Phi=-\frac{1}{2(d-1)}(t-\partial^2\tau)\ ;\qquad
2\phi-2\dot v+\ddot\tau
\ee
($\Phi$ is the Bardeen potential)
and the vector combination
$v^T_i-\dot\zeta_i$,
as well as $h^{TT}_{ij}$ are gauge invariant.
Also, the combination
$\dot t-2\partial^2 v$
is invariant under spacial diffeomorphisms ($\epsilon_0=0$).

We insert the new variables in (\ref{fierzpauliaction})
and calculate the Lagrangian $L=\int d^{d-1} x \cL$.
We allow integration by parts of spacial derivatives.
Also, we remove all time derivatives of $v^T_i$ and $v$ 
by adding suitable total time derivative terms. 
In this way we arrive at:
\bea
L&=&\int d^{d-1} x\Bigg[
\frac{1}{2}\left(\dot h^{TT}_{ij}\right)^2
-\dot\zeta_i\partial^2\dot\zeta_i
+2(d-1)(d-2)\dot \Phi^2
+2(d-2)\dot \Phi(\dot t-2\partial^2 v)
+2\dot\zeta_i \partial^2 v^T_i 
\nn &&
+\frac{1}{2}h^{TT}_{ij}\partial^2h^{TT}_{ij}
-v^T_i\partial^2 v^T_i
-2(d-2)(d-3)\Phi\partial^2\Phi
+4(d-2)\phi\partial^2\Phi
\Bigg]\ .
\label{FPLag}
\eea
We perform the Dirac constraint analysis on this Lagrangian.
From (\ref{FPLag}) we derive the conjugate momenta
\bea
\Pi^{TT}_{ij}&=&\dot h^{TT}_{ij}
\label{mTT}
\\
\Pi^{vT}_i&=&0
\label{mvT}
\\
\Pi^\zeta_i&=&-2\partial^2(\dot\zeta_i-v^T_i)
\label{mxi}
\\
\Pi^\phi&=&0
\label{mphi}
\\
\Pi^t&=&2(d-2)\dot\Phi
\label{mt}
\\
\Pi^v&=& 0
\label{mv}
\\
\Pi^\Phi&=&4(d-1)(d-2)\dot\Phi+2(d-2)(\dot t-2\partial^2 v)
\label{mPhi}
\eea
Equations \p{mTT}, \p{mxi}, \p{mt} and \p{mPhi} can be 
solved for the velocities
$\dot h^{TT}_{ij}$,
$\dot\zeta_i$,
$\dot\Phi$,
$\dot t$,
whereas \p{mvT}, \p{mphi}, \p{mv} give $d$ primary constraints.

Their preservation leads to $d$ secondary constraints
\be
\Pi^\zeta_i\ ,\qquad
\partial^2 \Phi\ ,\qquad
\partial^2 \Pi^t\ .
\label{scv}
\ee
There are no further constraints,
and all constraints are first class.
We then have to gauge fix the system. We can take the gauge fixing conditions
\be
v_i^T=0\ ,\qquad
\phi=0\ ,\qquad
v=0\ ,\qquad
\zeta_i=0\ ,\qquad
\Pi^\Phi=0\ ,\qquad
t=0.
\ee
The situation is very simple, because
each gauge condition is conjugate to one of the constraints.
If we order all the constraints as
$\phi_a=(\Pi^{vT}_i,\Pi^\phi,\Pi^v,\Pi^\zeta_i,
\partial^2\Phi,\partial^2\Pi^t)$
and all gauge conditions as
$\chi_a=(v_i^T,\phi,v,\zeta_i,\Pi^\Phi,t)$.
Then the matrix of Poisson brackets
$M_{ab}=\{\phi_a,\chi_b\}$
is diagonal and has determinant
\be
\det M=(\det(-\partial^2))^2\ .
\ee

The gauge conditions determine the Lagrange multipliers in the extended Hamiltonian, which in the chosen gauge becomes
\be
\label{hgf}
H_{GF}=\int d^{d-1}x\Bigg[
\frac{1}{2}\left(\Pi^{TT}_{ij}\right)^2
+\frac{1}{4}\Pi^\zeta_i\frac{1}{\partial^2}\Pi^\zeta_i
+\frac{d-1}{2(d-2)}
\left(\Pi^t\right)^2
-\frac{1}{2}h^{TT}_{ij}\partial^2h^{TT}_{ij}
-2(d-2)(d-3)\Phi\partial^2\Phi
\Bigg]
\ee
In particular on the constrained surface it is
\be
\label{constHamFP}
H_C=\int d^{d-1}x\Bigg[
\frac{1}{2}\left(\Pi^{TT}_{ij}\right)^2
+\frac{1}{2}\left(\partial_k h^{TT}_{ij}\right)^2
\Bigg]\ .
\ee

Let us now come to the path integral.
The measure is
$$
d\mu_{GR}=\cD h_{ij}^{TT}\, \cD \Pi_{ij}^{TT} \,
\cD v_i^T \,\cD \Pi^{vT}_i \,
\cD \zeta_i \,\cD \Pi^\zeta_i \,
\cD \phi \,\cD \Pi^\phi \,
\cD v \,\cD \Pi^v \,
\cD t \,\cD \Pi^t \,
\cD \Phi \,\cD \Pi^\Phi \,
$$
The Hamiltonian path integral is
\bea
Z_{GR}\hs{-2} &=&\hs{-2} \int d\mu_{FP}\,
\Pi_a\delta(\phi_a)
\,\Pi_b\delta(\chi_b)\det M
\nn
&& \hs{-7}
\times \exp \left\{i\int dt
\left[\int d^{d-1} x \left[ 
\dot h_{ij}^{TT} \Pi_{ij}^{TT} 
+ \dot v_i^T \Pi_{v_i^T}
+ \dot \zeta_i \Pi_{\zeta_i}
+\dot\phi\Pi^\phi
+ \dot v \Pi^v  
+ \dot t \Pi^t
+ \dot \Phi \Pi^\Phi\right] -  H 
\right]\right\}.
\nn
\label{pfpath}
\eea

Two of the secondary constraints contain $\partial^2$.
Using $\delta(ax)=(1/a)\delta(x)$, they give
$$
(\det(-\partial^2))^{-2}\delta(\Phi)\delta(\Pi^t)\ .
$$
This power of the determinant exactly cancels $\det M$
in the path integral.
All the variables are now integrated against a delta function,
except for the transverse traceless tensor and its momentum,
so the path integral reduces to
\bea
Z_{GR} \hs{-2}&=&
\hs{-2} \int \cD h_{ij}^{TT}\, \cD \Pi_{ij}^{TT}\,
 \exp\left\{ i\int dt\int d^{d-1} x \,
\left(\dot h_{ij}^{TT} \Pi_{ij}^{TT}- H_C\right) 
\right\}
\label{zgr}
\eea
where $H_C$ is given by (\ref{constHamFP}).
Finally integrating out the momentum
\be
Z_{GR}= \int \cD h_{ij}^{TT}\,
\exp\left\{\int dt L_C \right\}.
\ee
where
$$
L_C=\int d^{d-1}x\left[
\frac{1}{2}\left(\dot h^{TT}_{ij}\right)^2
-\frac{1}{2}\left(\partial_k h^{TT}_{ij}\right)^2
\right]\ .
$$
This is just the Lgrangian of the free fields 
$h^{TT}_{ij}$.
It can be written in the ``semi-covariant'' way
$$
L_C=\int d^{d-1}x\left[
-\frac{1}{2}\partial_\mu h^{TT}_{ij}
\partial^\mu h^{TT}_{ij}
\right]\ ,
$$
so that the Gaussian integral gives
\be
Z_{GR}=(\det\Box_{TT})^{-1/2}\ ,
\ee
where $\Box_{TT}$ is the d'Alembertian acting on $h^{TT}_{ij}$.
The number of independent components of a transverse traceless
tensor in $d-1$ space dimensions is $d(d-3)/2$, so
\be
Z_{GR}=(\det\Box)^{-d(d-3)/4}\ .
\label{zgrflat}
\ee

\subsection{Linearized minimal UG}

UG in minimal formulation corresponds to just setting $h=0$
in (\ref{fierzpauliaction}), which effectively just removes the last two terms.
In terms of the variables introduced in the preceding section,
the constraint $h=0$ implies $2\phi+t=0$.
Using this in (\ref{FPLag}), one finds
\bea
L_{UG}\!\!&=&\!\!\int d^{d-1}x\Bigg[
\frac{1}{2}\left(\dot h^{TT}_{ij}\right)^2
-\dot\zeta_i\partial^2\dot\zeta_i
+2(d-1)(d-2)\dot \Phi^2
+2(d-2)\dot \Phi\dot t
-4(d-2)\dot \Phi\partial^2 v
+2\dot\zeta_i \partial^2 v^T_i
\nonumber\\
&&
+\frac{1}{2}h^{TT}_{ij}\partial^2h^{TT}_{ij}
-v^T_i\partial^2 v^T_i
-2(d-2)(d-3)\Phi\partial^2\Phi
-2(d-2)\Phi\partial^2 t
\Bigg].~~~~
\label{UFPLag}
\eea
Compared to GR, we have one less scalar.
From (\ref{UFPLag}) we derive the conjugate momenta
\bea
\Pi^{TT}_{ij}&=&\dot h^{TT}_{ij}
\label{UmTT}
\\
\Pi^{vT}_i&=&0
\label{UmvT}
\\
\Pi^\zeta_i&=&-2\partial^2 (\dot\zeta_i- v^T_i)
\label{Umxi}
\\
\Pi^t&=&2(d-2)\dot\Phi
\label{Umt}
\\
\Pi^v&=&0
\label{Umv}
\\
\Pi^\Phi&=&4(d-1)(d-2)\dot\Phi+2(d-2)\dot t-4(d-2)\partial^2 v
\label{UmPhi}
\eea
Equations \p{UmTT}, \p{Umxi}, \p{Umt} and \p{UmPhi}
can be solved for the velocities $\dot h^{TT}_{ij}$,
$\dot\zeta_i$, $\dot\Phi$ and $\dot t$,
whereas \p{UmvT} and \p{Umv} give $d-1$ primary constraints
\be
C^{vT}_i=\Pi^{vT}_i\ ,\qquad
C^{v}=\Pi^v\ .
\label{Upcv}
\ee
Taking Poisson brackets
with the Hamiltonian one obtains $d-1$ secondary constraints
\be
S^{vT}_i=\Pi^\zeta_i\ ,\qquad
S^{v}= \pa^2\Pi^t\ .
\label{Uscv}
\ee
The first of these commutes with the Hamiltonian
and is therefore automatically conserved.
The second, however, generates a ``tertiary'' constraint
\be
T^v=\partial^2\partial^2\Phi\ ,
\ee
which is $\partial^2$ times the second constraint in (\ref{scv}).
Its Poisson bracket with the Hamiltonian is weakly zero,
so there are no further constraints.
All the constraints commute with each others 
and form a first class system.
Their total number is one less than in Fierz-Pauli theory.

We can impose the following gauge fixing conditions:
\bea
v_i^T=0\ ,\qquad
v=0\ ,\qquad
\zeta_i =0\ ,\qquad
t=0\ ,\qquad
\Pi^\Phi=0\ .
\eea

Let us order all the constraints as
$\phi_a=(C^{vT}_i,C^v,S^{vT}_i,S^v,T^v)$
and all gauge conditions as
$\chi_a=(v_i^T,v,\zeta_i,t,\Pi^\Phi)$.
Then the matrix of Poisson brackets
is the same as in GR except that the
rows for $\phi$ and $\Pi^\phi$ are missing,
and furthermore the term $T^v$-$\Pi^\Phi$ has an extra
factor $\partial^2$. Therefore
\be
\det M_{UG}=(\det(-\partial^2))^3\ .
\ee
The time conservation of the gauge constraints fixes
the Lagrange multipliers in the Hamiltonian.
When this is done, one finds that the gauge-fixed Hamiltonian
is again equal to (\ref{hgf}).

We finally come to the unimodular path integral.
The measure in the unimodular case is the same as for FP
except that $\phi$ and $\Pi^\phi$ are missing:
$$
d\mu_{UG}=\cD h_{ij}^{TT}\, \cD \Pi_{ij}^{TT} \,
\cD v_i^T \,\cD \Pi^{vT}_i \,
\cD \zeta_i \,\cD \Pi^\zeta_i \,
\cD v \,\cD \Pi^v \,
\cD t \,\cD \Pi^t \,
\cD \Phi \,\cD \Pi^\Phi \,
$$

The Hamiltonian path integral is
\bea
Z_{UG} &=& \int d\mu_{UG}\,
\Pi_a\delta(\phi_a)
\,\Pi_b\delta(\chi_b)\det M_{UG}
\nn
&& \times \exp \left\{i\int dt
\left[\int d^{d-1} x \left[ 
\dot h_{ij}^{TT} \Pi_{ij}^{TT} 
+ \dot v_i^T \Pi_{v_i^T}
+ \dot \zeta_i \Pi_{\zeta_i}
+ \dot v \Pi^v  
+ \dot t \Pi^t
+ \dot \Phi \Pi^\Phi\right] -  H 
\right]\right\}.
\nonumber
\eea

The constraints $S^v$ and $T^v$ contain $\partial^2$ and $\partial^2\partial^2$.
Using $\delta(ax)=(1/a)\delta(x)$, they become
$$
(\det(-\partial^2))^{-3}\delta(\Phi)\delta(\Pi^t)\ .
$$
Again, the power of the determinant exactly cancels $\det M_{UG}$
in the path integral.
All the variables are now integrated against a delta function,
except for the transverse traceless tensor and its momentum.
Thus, 
\be
Z_{UG}=Z_{GR}\ .
\label{zugflat}
\ee
In these ``unitary'' gauges, the two path integrals are the same.

\section{One loop effective actions}

The general framework for our calculations
is the same as in \cite{pereiraII}.
In view of application to UG we restrict ourselves to
the exponential parametrization (\ref{exppar}).

The symmetric tensor $h_{\mu\nu}$ is subjected to
the York decomposition
\begin{equation}
h_{\mu\nu}=h^{\mathrm{TT}}_{\mu\nu}
+\bar{\nabla}_{\mu}\xi_{\nu}+\bar{\nabla}_{\nu}\xi_{\mu}
+\left(\bar{\nabla}_{\mu}\bar{\nabla}_{\nu}
-\frac{1}{d}\bar{g}_{\mu\nu}\bar{\nabla}^{2}\right)\sigma
+\frac{1}{d}\bar{g}_{\mu\nu}h\,,
\label{ap1}
\end{equation}
where
\begin{equation}
\bar{\nabla}^{\mu}h^{\mathrm{TT}}_{\mu\nu}=0\,,\,\,\,\, \bar{g}^{\mu\nu}h^{\mathrm{TT}}_{\mu\nu}=0
\,,\,\,\,\, \bar{\nabla}^{\mu}\xi_{\mu}=0\,,\,\,\,\,
h=\bar{g}^{\mu\nu}h_{\mu\nu}\,.
\label{ap2}
\end{equation}
The Jacobian of the transformation
$h_{\mu\nu}\to\{h^{TT}_{\mu\nu},\xi_\mu,\sigma,h\}$
is
\be
\label{jac}
J_1=\det\left(\lich_1-\frac{2\bR}{d}\right)^{1/2}
\det\left(\lich_0\right)^{1/2}
\det\left(\lich_0-\frac{\bR}{d-1}\right)^{1/2}
\ee
where $\Delta_L$ are the Lichnerowicz Laplacians:
\bea
\lich_0 \phi &=& -\bnabla^2 \phi, \nn
\lich_1 A_\mu &=& -\bnabla^2 A_\mu + \bR_\mu{}^\rho A_\rho, \nn
\lich_2 h_{\mu\nu} &=& -\bnabla^2 h_{\mu\nu} 
+\bR_\mu{}^\rho h_{\rho\nu}
+ \bR_\nu{}^\rho h_{\mu\rho} 
-\bR_{\mu\rho\nu\s} h^{\rho\s} 
-\bR_{\mu\rho\nu\s} h^{\s\rho}\ .
\eea

Now consider an infinitesimal diffeomorphism $\epsilon^\mu$.
We can decompose the transformation
parameter $\epsilon^\mu$ in its longitudinal and transverse parts
(relative to the background metric):
\be
\label{qpar}
\epsilon^\mu=\epsilon^{T\mu}+\bnabla_\mu\phi\ ;
\qquad
\bar\nabla_\mu\epsilon^{T\mu}=0\ .
\ee
We can then calculate the separate transformation properties
of the York variables under longitudinal and transverse
infinitesimal diffeomorphisms.
We have
\be
\label{deltaxi}
\delta_{\epsilon^T}\xi^\mu=\epsilon^{T\mu}\ ;\qquad
\delta_\phi h=-2\lich_0\phi\ ;\qquad
\delta_\phi\sigma=2\phi\ ,
\ee
all other transformations being zero.
Note that $\sigma$ and $h$ are gauge-variant but the scalar combination
\be
\label{ess}
s=h+\lich_0\sigma
\ee
is invariant.

We will expand the action around an Einstein background
$\bR_{\mu\nu}=\frac{\bR}{d}\bg_{\mu\nu}$.
In GR, an Einstein metric automatically satisfies the tracefree part
of the Einstein equations with cosmological constant.
The remaining trace equation is $E=0$, where
\be
E=\bR-\frac{2d\Lambda}{d-2}\ .
\ee
In UG this equation is not present,
so an Einstein background is automatically on shell.

\subsection{One-loop GR}

Expanding the EH action around an Einstein background,
the Hessian is
(see \cite{pereiraII} or section 5.4.6 in \cite{perbook}):
\begin{eqnarray}
S\!\!&=&\!\!\frac{Z_N}{2}\!\int d^dx\,\sqrt{\bg}
\Bigg\{
\frac{1}{2}h^{\mathrm{TT}}_{\mu\nu}
\left(\Delta_{L2}-\frac{2\bR}{d}
\right)h^{\mathrm{TT}\mu\nu}
\nonumber\\
&&
\hskip2cm
-\frac{(d-1)(d-2)}{2d^2}s\left(\Delta_{L0}-\frac{\br}{d-1}\right)s
-\frac{d-2}{4d}E h^2
\Bigg\}\,.
\label{ehgf1}
\end{eqnarray}

A nice property of the exponential parametrization is that, 
aside from the term proportional to the EOM,
only the gauge-invariant variables $h^{TT}_{\mu\nu}$ and $s$
appear in the Hessian.

In the path integral, we add to the action a gauge-fixing term, typically of the form
\be
\label{gaugefixingterm}
S_{GF}=\frac{Z_N}{2\alpha}\int d^d x\sqrt{\bg}\,\bg^{\mu\nu}F_\mu F_\nu
\ee
with
\be
\label{gaugefixingcondition}
F_\mu=\bnabla_\rho h^\rho{}_\mu-\frac{\beta+1}{d}\bnabla_\mu h\ .
\ee

Using the York decomposition and defining
\be
\label{chi}
\chi=\frac{((d-1)\lich_0-\bR)\sigma+\beta h}{(d-1-\beta)\lich_0-\bR}\ ,
\ee
the gauge fixing condition reads
\be
\label{gfdecomposed2}
F_\mu=-\left(\lich_1-\frac{2\bR}{d}\right)\xi_\mu
-\frac{d-1-\beta}{d}\nabla_\mu
\left(\lich_0-\frac{\bR}{d-1-\beta}\right)\chi\ ,
\ee
where, using the Einstein condition,
$\lich_1=-\bnabla^2+\frac{\bR}{d}$.
The gauge fixing action is then equal to
\be
\label{gf2}
S_{GF}=\frac{Z_N}{2\alpha}\int d^d x\sqrt{\bg}
\Bigg[\xi_\mu\left(\lich_1-\frac{2\bR}{d}\right)^2\xi^{\mu}
+\frac{(d-1-\beta)^2}{d^2}
\chi\lich_0\left(\lich_0-\frac{\bR}{d-1-\beta}\right)^2\chi\Bigg]\ .
\ee
Under the transformation (\ref{deltaxi})
the variable $\chi$ transforms in the same way as $\sigma$.
Thus $\xi$ and $\chi$ can be viewed as the gauge degrees of freedom.

Decomposing the ghost into transverse and longitudinal parts
\be
C_\nu=C^T_\nu+\nabla_\nu\frac{1}{\sqrt{-\bnabla^2}}C^L
\label{ezio}
\ee
and likewise for $\bar C$,
the ghost action splits in two terms
\be
\label{ghostaction2}
S_{gh}=\int\! d^d x\sqrt{\bg}
\left[
\bar C^{T\mu}\!\left(\!\lich_1-\frac{2\bR}{d}\!\right)C^T_\mu
+2\frac{d-1-\beta}{d}
\bar C^L\!\left(\!\lich_0-\frac{\bR}{d-1-\beta}\!\right)C^L\right].
\ee
We note that the change of variables 
(\ref{ezio}) has unit Jacobian.

The (background-)gauge invariance of the Faddeev-Popov determinant
and of the classical action leads by the standard procedure
to the factorization of the volume of the gauge group
\be
V_\diff=\int (d\epsilon)\ .
\ee
This factor is usually dropped, but we will keep it 
explicit here for later reference.

The partition function is the product of determinants coming from
all the fields, ghosts and from the Jacobians.
Several determinants (among them all the scalar determinants, and in particular the $\beta$-dependent ones) cancel, 
and the partition function is independent of the gauge choice:
\be
Z_{GR}=V_\diff e^{-S(\bg)}
\frac
{\Det_1\left(\Delta_{L1}-\frac{2\br}{d}\right)^{1/2}}
{\Det_2\left(\Delta_{L2}-\frac{2\br}{d}\right)^{1/2}}\ .
\label{zgrc}
\ee

This formula explicitly agrees with (\ref{zgrflat})
when the background is flat:
aside from the field-independent prefactor,
the determinant on transverse vectors contributes
$\Det\Box^{(d-1)/2}$
while that on transverse tracefree tensors contributes
$\Det\Box^{-(d-2)(d+1)/4}$.
The powers of the determinants thus correctly
count the number of physical degrees freedom of the theory.

We note that in the limit $\beta\to\infty$,
$\chi\to\lich_0^{-1} h$ and the second term in (\ref{gf2})
imposes $h=0$ strongly.
This is equivalent to removing by hand $h$ from the
linearized action and adding a ghost term $\sqrt{\det\lich_0}$,
a procedure that was called ``unimodular gauge'' in \cite{pv1}.
This leaves a residual gauge freedom parametrized by the
transverse vecorfields $\epsilon^T_\mu$, that can be gauge-fixed
by setting $\xi_\mu=0$, or by adding a suitable gauge-fixing
term of a type that we shall discuss in the next section.

\subsection{One-loop UG in minimal formulation}

Let us now discuss UG in the minimal formulation.
We impose $h=0$ at the classical level, before
defining the functional integral.
Then we have to gauge fix $\sdiff$,
which is generated by vector fields satisfying
\be
\bnabla_\mu\epsilon^\mu=0\ .
\label{diffconstraint}
\ee
In order to define a suitable gauge-fixing for these transformations,
let
\be
L^\mu{}_\nu=\bnabla^\mu\frac{1}{\bnabla^2}\bnabla_\nu\ ;\qquad
T^\mu{}_\nu=\delta^\mu_\nu-L^\mu{}_\nu
\ee
be the longitudinal and transverse projectors defined
relative to the background metric. 
We choose our gauge-fixing function as
\footnote{It may be better to have a local gauge-fixing condition.
This can be achieved by inserting a power of $\lich_1$
in the gauge fixing term (\ref{uggf}) below,
see \cite{Eichhorn}.
Ultimately the additional determinant is canceled by
a Nielsen-kallosh ghost term, so that the final result
is the same.
In order to minimize the number of determinants we stick
to a non-local gauge fixing term.}
\be
F_\mu= T_{\mu\nu} \bnabla_\rho h^{\rho\nu}
=-\left(\lich_1-\frac{2\bR}{d}\right)\xi_\mu\ .
\label{uggaugecondition}
\ee

We can  follow the standard Faddeev-Popov procedure by inserting in
the functional integral the formal expression
\be
\label{one}
1=\int (d\epsilon^T)\Psi(h^{\epsilon^T},\bg)
\delta(F_\nu(h^{\epsilon^T},\bg))\ .
\ee
Passing to York variables and recalling that 
$\delta_{\epsilon^T}\xi_\nu=\epsilon^T_\nu$,
the evaluation of this expression leads to
$\Psi=\Det\left(\lich_1-\frac{2\bR}{d}\right)$.
As usual the delta function can be exponentiated as
the gauge-fixing term
\be
S_{GF}=\frac{Z_N}{2\alpha}\!\int d^dx\,\omega F_{\mu}T^{\mu\nu}F_{\nu}
=\frac{Z_N}{2\alpha}\!\int d^dx\,\omega
\,{\xi}_{\mu}
\left(-\bnabla^2-\frac{\bR}{d}
\right)^2{\xi}^{\mu}\ ,
\label{uggf}
\ee
while the Faddeev-Popov determinant $\Psi$ is exponentiated
as the ghost action
\be
S_{gh}=\int d^dx\,\omega
\,\bar C_{\mu}^T
\left(-\bnabla^2-\frac{\bR}{d}
\right)C^\mu{}^T \ ,
\label{ugghost}
\ee
where ghost and antighost fields are transverse vectors.

In the linearized Hilbert action, due to $h=0$, we can replace
$s$ by $\lich_0\sigma$.
The integration over $\sigma$ thus yields 
$\det\lich_0^{-1}\Det\left(\lich_0-\frac{\br}{d-1}\right)^{-1/2}$.
Finally we have to take into account the Jacobian (\ref{jac}).
The determinants work out as in GR, except that
one scalar determinant coming from the integration over $\sigma$
remains uncanceled because of the absence of the scalar ghost.
Collecting all the determinants, the partition function
of unimodular gravity is
\be
Z_{UG}=\left(\int (d\epsilon^T)\right) e^{-S(\bg)}
\frac
{\Det_1\left(\Delta_{L1}-\frac{2\br}{d}\right)^{1/2}}
{\Det_2\left(\Delta_{L2}-\frac{2\br}{d}\right)^{1/2}
\Det\lich_0^{1/2}}\ .
\label{zugc}
\ee

It would be tempting at this point to interpret the first
term on the r.h.s. as the volume of $\sdiff$ and to drop it
from the partition function.
One would then conclude that the partition function of UG differs
from that of GR by a scalar determinant.
This, however, cannot be correct.
It would mean that there is an additional scalar physical
degree of freedom, in contrast to the result
of the Hamiltonian analysis (\ref{zugflat}).

The key to a correct definition of the path integral is the
requirement that the volume of the gauge group, that one factors
and discards, has to be independent of the metric.
At the formal level of this discussion, all integrals 
over differentially unconstrained fields such as
$\int (d\epsilon)$ are metric-independent.
Since the definition of a transverse vector depends on the
metric, the integral $\int (d\epsilon^T)$ cannot be
treated as a metric-independent constant.
This suggests that the measure of the group $\sdiff$
must contain some determinant.
After all, we note that already the volume of $\diff$,
when written in terms of the
transverse and longitudinal gauge parameters,contains
a determinant. In fact from (\ref{qpar}) we find that
\be
(d\epsilon)=(d\epsilon^T)(d\phi)\Det\lich_0^{1/2}\ .
\label{anna}
\ee
One could define the volume of $\sdiff$ as the integral
$\int(d\epsilon)$ with a delta function 
$\delta(\nabla_\mu\epsilon^\mu)$.\footnote{We thank D. Benedetti for this remark.}
Using (\ref{anna}) and $\nabla_\mu\epsilon^\mu=-\lich_0\phi$,
we find that integrating a function that is independent of $\phi$
\bea
\int(d\epsilon)\delta(\nabla_\mu\epsilon^\mu)
&=&\int(d\epsilon^T)(d\phi)\Det\lich_0^{1/2}\delta(\lich_0\phi)
\nn
&=&\int(d\epsilon^T)(d\phi)\Det\lich_0^{-1/2}\delta(\phi)
=\int(d\epsilon^T)\Det\lich_0^{-1/2}\ .
\eea

This is the correct result, but from this argument 
the metric-independence is not evident. 
To this end, we proceed as follows.
We note that the gauge parameters $\epsilon^T$ and $\phi$ 
are the coordinates in the subgroup $\sdiff$ and in the
quotient space $Q=\diff/\sdiff$, respectively.
This quotient space can be identified with the
space of volume-forms.
Therefore we demand that the measure on the quotient space agrees
with the measure on the volume forms.
An infinitesimal change of volume form is a trace deformation
of the metric. 
Thus the measure on volume-forms is $(dh)$ and  $V_Q=\int(dh)$.
Equation (\ref{qpar}) implies that
$$
(dh)=(d\phi)\,\Det\lich_0\ .
$$
We can thus split (\ref{anna}) as follows
$$
V_\diff=\int (d\epsilon)=
\int(d\epsilon^T)\det\lich_0^{-1/2}
\int(d\phi)\det\lich_0
=V_\sdiff V_Q\ , 
$$
where
\be
V_\sdiff=\int (d\epsilon^T)\Det\lich_0^{-1/2}\ .
\ee
Now $V_\sdiff$ is seen to be the ratio
of $V_\diff$ and $V_Q$, both of which are integrals
over unconstrained variables and hence metric-independent.
Thus this definition of $V_\sdiff$ is metric-independent.\footnote{At this point one may wonder
whether in (\ref{one}) we should also have inserted $\det\lich_0^{-1/2}$.
This would have led to
a factor $\det\lich_0^{1/2}$ in $\Psi$,
and these two determinants would cancel out in the
path integral.}
This is also the expected result, because with this measure
we can rewrite (\ref{zugc}) as
\be
Z_{UG}=V_\sdiff e^{-S(\bg)}
\frac
{\Det_1\left(\Delta_{L1}-\frac{2\br}{d}\right)^{1/2}}
{\Det_2\left(\Delta_{L2}-\frac{2\br}{d}\right)^{1/2}}\ .
\label{zugc2}
\ee
Apart from the volume of the gauge group, that we can now drop,
the result is identical to the partition function for GR.

\subsection{One-loop UG with Weyl invariance}

We now consider the form (\ref{ester}) for the UG action.
The quantum field is now the unrestricted metric $\gamma_{\mu\nu}$
and in the background field expansion we have the option of using
either the exponential or the more traditional linear splitting.
Let us discuss first the case of the exponential parametrization.
Since the trace and tracefree parts of the fluctuation $h_{\mu\nu}$
commute, we can write
\be
\gamma_{\mu\nu}=e^{\frac{1}{d}h}\tilde g_{\mu\nu}\ ,\qquad
\tilde g_{\mu\nu}=\bg_{\mu\rho}\left(\exp(h^T)\right)^\rho{}_\nu
\ee
where $h^T$ is a traceless matrix and $h$ is a function.
The determinant of the metric $\gamma_{\mu\nu}$ is
\be
\det\gamma= e^h \det\tilde g=e^h \det\bg=e^h\omega^2\ .
\ee
Then,
$$
|\gamma|^{-1}\nabla|\gamma|=\nabla h+2\omega^{-1}\nabla\omega
$$
and the integrand of (\ref{ester}) becomes:
$$
\omega e^{\frac{1}{d}h}\left[R[\gamma]+(d-1)(d-2)(\partial \frac{1}{2d}h)^2\right]=\omega\,R[\tilde g]
$$
so that
\be
S(\gamma)=Z_N\int d^d x\, \omega\, R[\tilde g]\ .
\ee
It is now clear that in the expansion of the action (\ref{ester})
the trace $h$ cancels completely, and the rest 
is exactly as the expansion of the action (\ref{action}).
Without any further calculation, we conclude that
the effective action is the same.

In linear parametrization, a more detailed calculation is required.
Writing
\be
\gamma_{\mu\nu}=\bg_{\mu\nu}+h_{\mu\nu}
\ee
the Hessian of (\ref{ester}) is \cite{Alvarez:2006uu}
\bea
S\!\!&=&\!\!Z_N\!\int d^dx\,\sqrt{|\bg|}
\Bigg\{
\frac{1}{4}h_{\mu\nu}\bnabla^2 h^{\mu\nu}
-\frac{1}{2}h_{\mu\nu}\bnabla^\mu\bnabla_\rho h^{\rho\nu}
+\frac{1}{d}h\bnabla_\mu\bnabla_\nu h^{\mu\nu}
-\frac{d+2}{4d^2}h\bnabla^2 h
\nonumber
\\
&&+\frac{1}{2}h_{\mu\nu}\bR^{\mu\rho\nu\sigma}h_{\rho\sigma}
+\frac{1}{2}h_{\mu\nu}\bR^{\mu\rho}h_\rho{}^\nu
-\frac{1}{d}h\bR^{\rho\sigma}h_{\rho\sigma}
-\frac{1}{2d}\left(h_{\mu\nu}h^{\mu\nu}-\frac{1}{d}h^2\right)\bR
\Bigg\}\,.
\label{hess1}
\eea

For Euclidean signature,
assuming an Einstein background and rewriting
in terms of the Lichnerowicz Laplacians
\be
S\!\!=\!\!Z_N\!\int d^dx\,\sqrt{\bg}
\Bigg[
\frac{1}{4}h_{\mu\nu}\bDelta_2 h^{\mu\nu}
-\frac{1}{2}\bnabla^\mu h_{\mu\nu}\bnabla_\rho h^{\rho\nu}
-\frac{1}{d}h\bnabla_\mu\bnabla_\nu h^{\mu\nu}
-\frac{d+2}{4d^2}h\bDelta_0 h
-\frac{1}{2d}\bR\left(h_{\mu\nu}^2-\frac{1}{d} h^2\right)
\Bigg]
\label{hess2}
\ee
This Hessian has a kernel consisting of infinitesimal diffeos
and infinitesimal Weyl transformations.
We can then fix the gauge $h=0$ for the Weyl group.
This leaves no ghosts, because $h$ transforms by a shift.
Using the York decomposition for the remaining traceless
fluctuation leads back exactly to (\ref{ehgf1}).
For $\sdiff$ we can fix the same gauge as in section 2.1,
so that the result for the effective action is again the same.

\subsection{Four dimensions}

The effective action is related to the partition function by $Z(\bg)=e^{-\Gamma(\bg)}$.
Neglecting field-independent terms, we find indifferently from
(\ref{zgrc}) or (\ref{zugc2}):
\be
\Gamma(\bg)=S(\bg)
+\frac{1}{2}\log\Det\left(\Delta_{L2}-\frac{2\br}{d}\right)
-\frac{1}{2}\log\Det\left(\Delta_{L1}-\frac{2\br}{d}\right)\ .
\label{divea}
\ee
The divergent part of the effective action
can be computed by standard heat kernel methods \cite{perbook}.
On an Einstein background in four dimension the 
logarithmically divergent part is
\bea
\Gamma_{log}(\bg)&=& -\frac{1}{2(4\pi)^2}
\int d^4x\,\sqrt{\bg}
\log\left(\frac{k^2}{\mu^2}\right)\left(
\frac{53}{45}\bR_{\mu\nu\rho\sigma}\bR^{\mu\nu\rho\sigma}
-\frac{29}{40}R^2
\right)
\,,
\label{gammaabc}
\eea
where $k$ stands for a cutoff
and we introduced a reference mass scale $\mu$.
Replacing $R\to4\Lambda$ this agrees with the
classic result of \cite{Christensen:1979iy}.

The result is independent of the gauge-fixing parameters $\alpha$
and $\beta$ as expected in view of the fact that 
the Einstein condition 
is enough to put the background metric on shell.

\section{Discussion}

UG and GR are known to be almost identical at the classical level.
Besides the equations of motion, this has been seen also
in the tree-level amplitudes
\cite{Burger:2015kie,Alvarez:2016uog}.
The question is then whether this persists when loop effects
are taken into account.
In this paper we provide further evidence for the equivalence
of the two theories.
The main subtlety arises in the definition of the path integral 
for UG. 
Since this is tied closely to implementing the proper 
count of degrees of freedom,
let us review quickly how this works, 
starting from the classical theory.

In GR, as in Yang-Mils theory, the number of physical 
degrees of freedom is equal to the number of fields minus
twice the number of gauge parameters
(``the gauge strikes twice'').
For example, in four-dimensional gravity we have 10 fields
$h_{\mu\nu}$ and four gauge parameters (the components of
a vector field $\epsilon$) yielding two 
propagating degrees of freedom.
In UG we have nine components for $h_{\mu\nu}$ (which is tracefree)
and three gauge parameters
(the components of a transverse vector field $\epsilon^T$)
so the general rule would seem to give three propagating
degrees of freedom.
The general rule does not apply in this case because the
gauge parameter is subject to the differential constraint
(\ref{diffconstraint}).

For a correct counting it is best to go back 
to the Hamiltonian formulation.
GR has 10 Lagrangian variables and 4 Lagrangian gauge parameters,
equivalently 20 Hamiltonian variables and 8 first class constraints
(i.e. 8 gauge parameters).
Each of these removes two Hamiltonian variables: one is removed by the
constraint itself and one by the corresponding gauge condition.
Thus one is left with $20-2\times 8=4$ canonical variables, 
corresponding to the 2 physical polarization states of the graviton.
In UG there are 9 fields and 3 Lagrangian gauge parameters,
giving rise to 6 constraints, but there is also a ``tertiary''
constraint bringing the number of first class constraints to 7.
With 18 Hamiltonian variables and 7 first class constraints,
we have again $18-2\times7=4$ physical canonical variables.
Using ``unitary'' gauges, all this can be easily implemented in the
Hamiltonian path integrals, showing that they are the same
for GR and UG. The price one has to pay for this simple count is the
lack of explicit covariance.


In the classical Lagrangian formulation of UG, when we use York variables,
the gauge condition (\ref{uggaugecondition}) removes $\xi_\mu$
and there seems to be a leftover unphysical 
scalar degree of freedom $\sigma$.
However, among the generators of $\sdiff$, there is a subclass
that is generated by vector fields that are, so to speak,
``simultaneously longitudinal and transverse'':
these are the vector fields of the form $\epsilon_\mu=\nabla_\mu\phi$
with $\lich_0\phi=0$.\footnote{These transformations are analogous to the
residual gauge transformation satisfying $\partial^2\epsilon_\mu=0$,
when one imposes the de Donder condition.}
The scalar $\sigma$ transforms by a shift and 
when it is on shell it can be removed by such 
a transformation, restoring the correct counting.\footnote{We note that the count of degrees
of freedom has to be done in Lorentzian signature, 
where the kernel of $\lich_0$ is an
infinite-dimensional space, parametrized by all the fields 
$\phi$ on an initial spacelike hypersurface.
On a compact Euclidean manifold without boundary
the kernel of $\lich_0$ consists only of the constants.}

In the path integral this issue manifests itself as
a nontrivial scalar determinant left over by the integration
on the field $\sigma$ 
(see the denominator of (\ref{zugc})).
We have shown that this determinant is part of the
measure on the group $\sdiff$ and therefore does not appear
in the ``physical'' terms in the final formula for the partition function.

We have framed this discussion in the context of a one-loop
definition of the path integral.
However, the main point concerned the definition of the
path integral measure.
Insofar as the path integrals embody all quantum effects,
they are the same for GR and UG.
This is in accordance with the conclusions of 
\cite{Fiol:2008vk,Padilla:2014yea}
for the perturbative expansion, and \cite{Bufalo:2015wda}
for the fully diffeomorphism-invariant version of UG.
Our results also agree with the result of \cite{Smolin}
that the effective action of UG is still a functional
of an unimodular metric.
One could try to make a non-perturbative statement
by using the functional renormalization group equations.\footnote{In particular it would be interesting
to compare the well-established Effective Average Action formalism \cite{Reuter1}
to a proper-time flow equation \cite{Floreanini}
whose exact version for the Wilsonian action
has been discussed recently \cite{deAlwis}.}

The preceding discussion was entirely in the context of the
``minimal'' formulation based on the exponential parametrization,
where GR is the quantum theory of a symmetric tensor
and UG is the quantum theory of a traceless symmetric tensor.
We have shown that the equivalence also holds in the
formulation in which the metric is unconstrained but there
is an additional Weyl symmetry.
In this formulation, a scalar degree of freedom is removed by 
imposing a Weyl gauge condition and 
the equivalence between UG and GR
at the quantum level hinges on the absence of Weyl anomalies.
In this connection we observe that if we define a scalar field 
$$
\phi^2
=\frac{\omega^{2/d}}{|\gamma|^{1/d}}
$$
then equation (\ref{ester}) becomes the usual action
for a conformally coupled scalar.
The absence of Weyl anomalies in such theories has
been proven in \cite{Percacci:2011uf,Codello:2012sn},
see also \cite{Alvarez:2013fs,Carballo-Rubio:2015kaa}.

With regards to the question left open in \cite{pereiraI},
whether the limit $m\to-1/d$ is continuous,
we find that the answer is positive.

Concerning the problem of the cosmological constant,
the classical conclusion that
vacuum energy does not gravitate extends to the quantum theory.
As already noted in \cite{Smolin},
this follows from the fact that the effective action will also
be ``unimodular'', so that the cosmological term
appears in the equations as an arbitrary integration constant.
This holds independently whether the quantum theory
is viewed merely as an effective field theory
or has an UV completion as in asymptotic safety.
The quantum treatment would be relevant
even if it turned out that gravity is ``emergent''
\cite{Percacci:2010af}.
A unimodular effective action 
eliminates the ``prediction'' that spacetime should
have Planckian curvature, but it does not explain why
it has the observed value, if we assume that the observed
cosmic acceleration is due to a cosmological term.

\section*{Acknowledgment}

We would like to thank Dario Benedetti, Astrid Eichhorn and  Taichiro Kugo for valuable discussions.
This work was supported in part by the Grant-in-Aid for Scientific Research Fund of the JSPS (C) No. 16K05331.
One of the authors (N.O.) would like to thank SISSA for kind hospitality during his visit when this work was started.



\begin{thebibliography}{99}


\bibitem{Anderson:1971pn}
J.~L.~Anderson and D.~Finkelstein,
``Cosmological constant and fundamental length,''
Am.\ J.\ Phys.\  {\bf 39} (1971) 901.

\bibitem{vanderBij:1981ym}
J.~J.~van der Bij, H.~van Dam and Y.~J.~Ng,
``The Exchange of Massless Spin Two Particles,''
Physica {\bf 116A} (1982) 307.

\bibitem{Buchmuller:1988wx}
W.~Buchmuller and N.~Dragon,
``Einstein Gravity From Restricted Coordinate Invariance,''
Phys.\ Lett.\ B {\bf 207} (1988) 292.\\
%
``Gauge Fixing and the Cosmological Constant,''
Phys.\ Lett.\ B {\bf 223} (1989) 313.

\bibitem{Ellis:2010uc}
G.~F.~R.~Ellis, H.~van Elst, J.~Murugan and J.~P.~Uzan,
``On the Trace-Free Einstein Equations as a Viable Alternative to General Relativity,''
Class.\ Quant.\ Grav.\  {\bf 28} (2011) 225007
[arXiv:1008.1196 [gr-qc]].
\\
G.~F.~R.~Ellis,
``The Trace-Free Einstein Equations and inflation,''
Gen.\ Rel.\ Grav.\  {\bf 46} (2014) 1619
[arXiv:1306.3021 [gr-qc]].

\bibitem{Henneaux:1989zc}
M.~Henneaux and C.~Teitelboim,
``The Cosmological Constant and General Covariance,''
Phys.\ Lett.\ B {\bf 222} (1989) 195.\\
%
M.~Henneaux, C.~Teitelboim and J.~Zanelli,
``Gauge Invariance and Degree of Freedom Count,''
Nucl.\ Phys.\ B {\bf 332} (1990) 169.

\bibitem{Smolin}
L.~Smolin,
``The Quantization of unimodular gravity and the cosmological constant problems,''
Phys.\ Rev.\ D {\bf 80} (2009) 084003
[arXiv:0904.4841 [hep-th]].

\bibitem{Padilla:2014yea}
A.~Padilla and I.~D.~Saltas,
``A note on classical and quantum unimodular gravity,''
Eur.\ Phys.\ J.\ C {\bf 75} (2015) no.11,  561
[arXiv:1409.3573 [gr-qc]].

\bibitem{Bufalo:2015wda}
R.~Bufalo, M.~Oksanen and A.~Tureanu,
``How unimodular gravity theories differ from general relativity at quantum level,''
Eur.\ Phys.\ J.\ C {\bf 75} (2015) 477
[arXiv:1505.04978 [hep-th]].

\bibitem{Fiol:2008vk}
B.~Fiol and J.~Garriga,
``Semiclassical Unimodular Gravity,''
JCAP {\bf 1008} (2010) 015
[arXiv:0809.1371 [hep-th]].
  
\bibitem{Alvarez}
E.~\'Alvarez, S.~Gonz\'alez-Mart\'\i n, M.~Herrero-Valea and C.~P.~Mart\'\i n,
``Unimodular Gravity Redux,''
Phys.\ Rev.\ D {\bf 92} (2015) no.6,  061502
[arXiv:1505.00022 [hep-th]].\\
%
``Quantum Corrections to Unimodular Gravity,''
JHEP {\bf 1508} (2015) 078
[arXiv:1505.01995 [hep-th]].

\bibitem{Eichhorn}
A.~Eichhorn,
``On unimodular quantum gravity,''
Class.\ Quant.\ Grav.\  {\bf 30} (2013) 115016
[arXiv:1301.0879 [gr-qc]];\\
``The Renormalization Group flow of unimodular $f(R)$ gravity,''
JHEP {\bf 1504} (2015) 096
[arXiv:1501.05848 [gr-qc]].

\bibitem{Benedetti:2015zsw}
D.~Benedetti,
``Essential nature of Newton’s constant in unimodular gravity,''
Gen.\ Rel.\ Grav.\  {\bf 48} (2016) no.5,  68
[arXiv:1511.06560 [hep-th]].

\bibitem{Saltas:2014cta}
I.~D.~Saltas,
``UV structure of quantum unimodular gravity,''
Phys.\ Rev.\ D {\bf 90} (2014) no.12,  124052
[arXiv:1410.6163 [hep-th]].

\bibitem{Alvarez:2006uu}
E.~\'Alvarez, D.~Blas, J.~Garriga and E.~Verdaguer,
``Transverse Fierz-Pauli symmetry,''
Nucl.\ Phys.\ B {\bf 756} (2006) 148
[hep-th/0606019].

\bibitem{Blas:2011ac}
D.~Blas, M.~Shaposhnikov and D.~Zenhausern,
``Scale-invariant alternatives to general relativity,''
Phys.\ Rev.\ D {\bf 84} (2011) 044001
[arXiv:1104.1392 [hep-th]]. \\ 
%
G.~K.~Karananas and M.~Shaposhnikov,
``Scale invariant alternatives to general relativity. II. Dilaton properties,''
Phys.\ Rev.\ D {\bf 93} (2016) no.8,  084052
[arXiv:1603.01274 [hep-th]].

\bibitem{Bonifacio:2015rea}
J.~Bonifacio, P.~G.~Ferreira and K.~Hinterbichler,
``Transverse diffeomorphism and Weyl invariant massive spin 2: Linear theory,''
Phys.\ Rev.\ D {\bf 91} (2015) 125008
[arXiv:1501.03159 [hep-th]].
  
\bibitem{Kawai:1989yh}
H.~Kawai and M.~Ninomiya,
``Renormalization Group and Quantum Gravity,''
Nucl.\ Phys.\  B {\bf 336} (1990) 115;
\\
H.~Kawai, Y.~Kitazawa and M.~Ninomiya,
``Ultraviolet stable fixed point and scaling relations in (2+epsilon)-dimensional quantum gravity,''
Nucl.\ Phys.\ B {\bf 404} (1993) 684
[hep-th/9303123].
\\
T.~Aida, Y.~Kitazawa, J.~Nishimura and A.~Tsuchiya,
``Two loop renormalization in quantum gravity near two-dimensions,''
Nucl.\ Phys.\  B {\bf 444} (1995) 353
[hep-th/9501056].

\bibitem{pv1}
R.~Percacci and G.~P.~Vacca,
``Search of scaling solutions in scalar-tensor gravity,''
Eur.\ Phys.\ J.\ C {\bf 75} (2015) 188
[arXiv:1501.00888 [hep-th]].

\bibitem{nink}
A.~Nink,
``Field Parametrization Dependence in Asymptotically Safe Quantum Gravity,''
Phys.\ Rev.\ D {\bf 91} (2015) 044030
[arXiv:1410.7816 [hep-th]]. \\
M.~Demmel and A.~Nink,
Phys.\ Rev.\ D {\bf 92} (2015) 104013
[arXiv:1506.03809 [gr-qc]].
\\
A.~Nink and M.~Reuter,
``The unitary conformal field theory behind 2D Asymptotic Safety,''
JHEP {\bf 1602} (2016) 167
[arXiv:1512.06805 [hep-th]].

\bibitem{pereiraI}
N.~Ohta, R.~Percacci and A.~D.~Pereira,
``Gauges and functional measures in quantum gravity I: Einstein theory,''
JHEP {\bf 1606} (2016) 115
[arXiv:1605.00454 [hep-th]].

\bibitem{pereiraII}
N.~Ohta, R.~Percacci and A.~D.~Pereira,
``Gauges and functional measures in quantum gravity II: Higher derivative gravity,''
arXiv:1610.07991 [hep-th].

\bibitem{baaklini}
N.~S.~Baaklini, M.~Tuite,
``Dirac Quantization of Spin-2 Field,''
J.\ Phys.\ A {\bf 12} (1979) L13.

\bibitem{perbook}
R. Percacci `` An introduction to covariant quantum gravity and asymptotic safety'',
World Scientific, Singapore (2017).

\bibitem{Christensen:1979iy}
S.~M.~Christensen and M.~J.~Duff,
``Quantizing Gravity with a Cosmological Constant,''
Nucl.\ Phys.\ B {\bf 170} (1980) 480.

\bibitem{Burger:2015kie}
D.~J.~Burger, G.~F.~R.~Ellis, J.~Murugan and A.~Weltman,
``The KLT relations in unimodular gravity,''
arXiv:1511.08517 [hep-th].  

\bibitem{Alvarez:2016uog}
E.~\'Alvarez, S.~Gonzalez-Martin and C.~P.~Martin,
``Unimodular Trees versus Einstein Trees,''
Eur.\ Phys.\ J.\ C {\bf 76} (2016) no.10,  554
[arXiv:1605.02667 [hep-th]].

\bibitem{Reuter1}
M.~Reuter,
``Nonperturbative evolution equation for quantum gravity,''
Phys.\ Rev.\ D {\bf 57} (1998) 971
[hep-th/9605030].
  
\bibitem{Floreanini}
R.~Floreanini and R.~Percacci,
``The Heat kernel and the average effective potential,''
Phys.\ Lett.\ B {\bf 356} (1995) 205
[hep-th/9505172].

\bibitem{deAlwis}
S.~P.~de Alwis,
``Exact RG Flow Equations and Quantum Gravity,''
arXiv:1707.09298 [hep-th].
  
\bibitem{Percacci:2010af}
R.~Percacci and G.~P.~Vacca,
``Asymptotic Safety, Emergence and Minimal Length,''
Class.\ Quant.\ Grav.\  {\bf 27} (2010) 245026
[arXiv:1008.3621 [hep-th]].  

\bibitem{Percacci:2011uf}
R.~Percacci,
``Renormalization group flow of Weyl invariant dilaton gravity,''
New J.\ Phys.\  {\bf 13} (2011) 125013
[arXiv:1110.6758 [hep-th]].

\bibitem{Codello:2012sn}
A.~Codello, G.~D'Odorico, C.~Pagani and R.~Percacci,
``The Renormalization Group and Weyl-invariance,''
Class.\ Quant.\ Grav.\  {\bf 30} (2013) 115015
[arXiv:1210.3284 [hep-th]].

\bibitem{Alvarez:2013fs}
E.~\'Alvarez and M.~Herrero-Valea,
``No Conformal Anomaly in Unimodular Gravity,''
Phys.\ Rev.\ D {\bf 87} (2013) 084054
[arXiv:1301.5130 [hep-th]].

\bibitem{Carballo-Rubio:2015kaa}
R.~Carballo-Rubio,
``Longitudinal diffeomorphisms obstruct the protection of vacuum energy,''
Phys.\ Rev.\ D {\bf 91} (2015) no.12,  124071
[arXiv:1502.05278 [gr-qc]].

\end{thebibliography}
\end{document}